\documentclass[12pt]{iopart}

\usepackage{iopams} 
\usepackage{setstack}
\usepackage{graphicx}
\usepackage{xcolor}
\usepackage{tikz}

\begin{document}

\title{Integrability of two-species partially asymmetric exclusion processes}

\author{Ivan Lobaskin$^1$, Martin R Evans$^1$ and Kirone Mallick$^2$}

\address{$^1$ School of Physics and Astronomy, University of Edinburgh, Peter Guthrie Tait Road, Edinburgh, EH9 3FD, United Kingdom}
\address{$^2$ Institut de Physique Th{\' e}orique, Universit{\' e} Paris-Saclay, CEA and CNRS, 91191 Gif-sur-Yvette, France}
\eads{\mailto{ivan.lobaskin@ed.ac.uk}, \mailto{m.evans@ed.ac.uk}, \mailto{kirone.mallick@cea.fr}}
\vspace{10pt}
\begin{indented}
\item[] \today
\end{indented}

\begin{abstract}
{We work towards the classification of} all one-dimensional exclusion processes with two species of particles that can be solved by a nested coordinate Bethe Ansatz. 
Using the Yang-Baxter equations, we obtain conditions on the model parameters that ensure that the underlying system is integrable. 
Three classes of integrable models are thus found.   
Of these, two classes are well known in literature, but the third has not been studied until recently, and never in the context of the Bethe ansatz.  
The Bethe equations are derived for the latter model as well as for the associated dynamics encoding the large deviation of the currents.
\end{abstract}

\section{Introduction}
In statistical physics far from equilibrium, the one-dimensional asymmetric simple exclusion process (ASEP) is a paradigmatic model, playing a similar role to the Ising model for equilibrium systems \cite{derrida2007non,blythe2007nonequilibrium,chou2011non}.
Its simplicity means that it can be used as a generic model in many fields, including biophysics \cite{macdonald1968kinetics,SNCM2018,scott2019power}, traffic modelling \cite{wolf1996traffic} and microfluidics \cite{cividini2017driven}. 
At the same time, a multitude of exact results have been derived, which is rare for interacting $N$-body problems.
The two main approaches for obtaining exact results have been the matrix product ansatz \cite{dehp1993} and the Bethe ansatz (BA) \cite{Alexander,Dhar,gwa1992bethe}.
For general reviews of these methods in the ASEP context, see \cite{blythe2007nonequilibrium} for the matrix product approach (which will not be used  in the present work) and \cite{golinelli2006asymmetric} for the BA.

The BA was first developed to diagonalize one-dimensional quantum spin chain Hamiltonians \cite{bethe1931theorie}, but has since been extended to many other fields, such as 2D vertex models \cite{baxter2016exactly} and ASEPs.
It was first shown that it can be applied to the  exclusion process on a ring \cite{Alexander,Dhar,gwa1992bethe}.
Since then, it has also been used on systems with various boundary conditions \cite{schutz1997exact,sasamoto1998one,de2005bethe,simon2009construction}, partially asymmetric (PASEP) models \cite{Kim,alcaraz1999exact,alcaraz2000exact,de2008slowest,prolhac2008current,prolhac2010tree} and multi-species cases \cite{derrida1999bethe,alcaraz1999exact,alcaraz2000exact,cantini2008algebraic}.
Using a deformation of the Markov operator, it was first shown in \cite{derrida1998exact} that the BA can be used very effectively to derive higher order particle displacement statistics, such as the diffusion coefficient.
For some particularly simple cases, it has even been possible to calculate the full statistics to all orders \cite{derrida1998exact,prolhac2010tree}.

The question of which systems can be solved by the BA, i.e. are integrable,  has a long history in the quantum and classical equilibrium contexts \cite{jimbo1990yang}. 
An answer to this problem is through the Yang-Baxter equations (YBE), which can be interpreted as consistency conditions for the BA; indeed, by checking whether the YBE hold on a general class of models, it is possible to select values of the  parameters  for which  exact solutions exist. 
Though this approach was  used for quantum \cite{kulish1982solutions} and vertex model  \cite{vieira2021solutions}, it has been applied less systematically to  study classical interacting particles systems, see for example  \cite{Schutz,cantini2008algebraic,Crampe2,crampe2014integrable}.

In this paper, we  conduct such a systematic study for two-species exclusion processes on a ring.
By postulating a very general coordinate BA and deriving the YBE, we are able to assess the integrability of a wide range of models, which suggests a classification of integrable two-species systems.

 We find that there are three families of integrable models: two of them are well-known in the literature, but there exists another, special, model that has only come to attention recently \cite{lobaskin2022matrix} and has not yet been studied from the viewpoint of integrability.
There, its steady state was solved using the matrix product approach, and the phase diagram, currents and density profiles were calculated.
This model has the peculiar feature that it is only integrable when there is only one particle of one of the species, which can be seen clearly in the YBE.
We conjecture that our study entirely exhausts all integrable two-species exclusion processes.

The remainder of the paper is structured as follows.
In \sref{BAderivation}, we show how the time evolution problem of a two-species PASEP can be solved by a general form of the coordinate BA. In addition, we show that  a very similar ansatz can be used to solve the conditioned (or deformed) time evolution problem, which allows direct calculation of the large deviations of the current.
In \sref{YBE}, we derive the YBE  and present  the three classes of solutions.
Two of these are well-known in literature \cite{derrida1993exact,mallick1996shocks,derrida1998exact,alcaraz1999exact,cantini2008algebraic}, but the third has only come to attention recently \cite{lobaskin2022matrix} and has not yet been studied in the context of the BA.
In \sref{newmodelBA}, we focus on this new model and derive the Bethe equations for the deformed time evolution problem.
In \ref{ybexample}, we explain the procedure that we used to find the solutions of the YBE.
Finally, in \ref{beqalgebra}, we fill in some algebraic steps in the calculation of \sref{newmodelBA}.

\section{Bethe ansatz solution of two-species PASEP problem} \label{BAderivation}
\subsection{Problem statement}
We consider an $L$ site ring with particles of two-species hopping stochastically in continuous time.
Let there be $M_1$ particles of species $1$ and $M_2$ particles of species $2$, with $M_1+M_2=M$.
For the most general two-species partially asymmetric problem with overtaking, the following processes take place with the given rates:
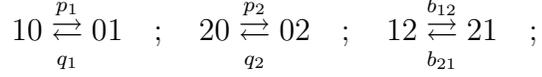
\begin{eqnarray*}
    10
    \underset{q_1}{\overset{p_1}{\rightleftarrows}}
    01 \quad ; \quad
    20
    \underset{q_2}{\overset{p_2}{\rightleftarrows}}
    02 \quad ; \quad
    12
    \underset{b _{21}}{\overset{b _{12}}{\rightleftarrows}}
    21 \quad ;    
\end{eqnarray*}
\newline
where $0,1,2$ represent empty sites, and particles of species $1$ and $2$, respectively.

Fixing the reference frame, we denote by $Q_j$ and $x_j$ respectively the species and position of particle $j$, counting from the left, where $Q_j\in \{1,2\}$ and $x_j \in {\mathbb Z} _L$.
As a shorthand, we denote the configuration of the system as ${\cal C} = \{ Q_1 \dots Q_M ; x_1 \dots x_M \}$.
The time evolution of the probability of a configuration ${\cal C}$ at time $t$ is given by the master equation
\begin{equation}
    \frac{\partial P_t ({\cal C})}{\partial t} = 
    \sum\limits_{{\cal C}'} {\cal M}({\cal C},{\cal C}') P_t({\cal C}') \;,
    \label{master}
\end{equation}
where $\cal M$ is the continuous-time  Markov operator containing the transition rates from configuration ${\cal C}'$ to ${\cal C}$.
We wish to solve the eigenvalue-eigenvector problem for the Markov operator.
It can be written in the form
\begin{eqnarray}
    \fl
    \lambda \psi ^{Q_1 \dots Q_M} (x_1 \dots x_M) = 
    -[M_1 (p_1+q_1) + M_2(p_2+q_2)]\psi ^{Q_1 \dots Q_M} (x_1\dots x_M) \nonumber
    \\
    \quad +\sum\limits_{j=1}^{M} [p_{Q_j} \psi ^{Q_1 \dots Q_M} (\dots  x_j -1\dots  )
    +q_{Q_j} \psi ^{Q_1 \dots Q_M} (\dots  x_j +1\dots )] \;,
    \label{evproblem}
\end{eqnarray}
where $\lambda$ is an eigenvalue and $\psi$ is the corresponding eigenvector.

This equation holds in configurations in which no two particles are on adjacent sites.
It can be assumed to hold in all configurations, provided an additional set of equations, called collision equations, also hold.
Considering the $j$-th and $(j+1)$-th particles, with $x_{j+1}=x_{j}+1$, these are given by the following, where $Q_{j,j+1}=1,2$,
\begin{eqnarray}
    \fl -(p_{Q_j} + q_{Q_{j+1}})\psi^{\dots Q_jQ_{j+1}\dots}(\dots x_j,x_j+1 \dots) 
    + p_{Q_{j+1}} \psi^{\dots Q_jQ_{j+1}\dots}(\dots x_j,x_j \dots) \nonumber \\ 
    \fl \qquad + q_{Q_j} \psi^{\dots Q_jQ_{j+1}\dots}(\dots x_j+1,x_j+1 \dots) 
    = -b_{Q_j Q_{j+1}}\psi^{\dots Q_jQ_{j+1}\dots}(\dots x_j,x_j+1 \dots)\nonumber \\
    \fl \qquad + b_{Q_{j+1}Q_j}\psi^{\dots Q_{j+1}Q_{j}\dots}(\dots x_j,x_j+1 \dots)\;,
    \label{collpsi}
\end{eqnarray}
where $b_{22}=b_{11}=0$.

Finally, the periodic boundary conditions are imposed via the requirement
\begin{equation}
    \psi ^{Q_1 Q_2 \dots Q_M} (x_1, x_2 \dots x_M) = 
    \psi ^{Q_2\dots Q_M Q_1} (x_2\dots x_M, x_1+L) \;. \label{pbc}
\end{equation}

Then  equations \eref{evproblem}, \eref{collpsi}, \eref{pbc} define the eigenvalue-eigenvector problem to be solved.

\subsection{Bethe ansatz solution}\label{BAnormal}
We use the following modified nested coordinate BA,
\begin{equation}
    \psi ^{Q_1 \dots Q_M} (x_1 \dots x_M) =
    r_{Q_1} ^{x_1} \dots r_{Q_M} ^{x_M} \sum\limits _\sigma A^{Q_1 \dots Q_M}_\sigma
    z_{\sigma (1)}^{x_1} \dots  z_{\sigma (M)}^{x_M} \;,
    \label{ansatz}
\end{equation}
where $\sigma $ indicates the permutations of the integers $1, \dots ,M$; the $A$'s and $z$'s are undetermined complex amplitudes and Bethe roots respectively; and the $r$'s are species dependent pre-factors, to be determined.
This is the most general known coordinate ansatz and is a straightforward nested generalization of the ansatz introduced in \cite{derrida1999bethe}.
There, it was used for a totally asymmetric problem with one defect particle.

The ansatz \eref{ansatz} is found to solve \eref{evproblem} provided the following condition holds:
\begin{equation}
    p_1q_1=p_2q_2\;. \label{pqpq}
\end{equation}
In this case, we can set
\begin{equation}
    p_1 = \alpha p_2\;,\qquad q_1=q_2/\alpha \;, \label{alpha}
\end{equation}
for some constant $\alpha $.
Then the ansatz \eref{ansatz} works, with
\begin{equation}
    r_1 = \alpha \;,\qquad r_2 =1\;.\label{r1r2}
\end{equation}
We assume $\alpha\neq0$, as otherwise the ansatz vanishes.
Notably, this excludes the model studied by Arndt \etal \cite{arndt1998spontaneous,rajewsky2000spatial}, which can be defined with our notation as $p_1=0,q_2=0$.
However, as this is believed to be non-integrable \cite{personal}, this is not of major concern.

Putting \eref{ansatz} with \eref{alpha} into \eref{evproblem} gives the following expression for the eigenvalue in terms of the Bethe roots,
\begin{equation}
    \lambda = -[M_1 (\alpha p_2+q_2/\alpha ) + M_2(p_2+q_2)] + \sum\limits_{j=1}^{M}( p_2z_j ^{-1} + q_2z_j )\;. \label{lambda}
\end{equation}

We remark that condition \eref{pqpq} comes from the form of the ansatz \eref{ansatz}.
However, a similar condition was derived from the the algebraic Bethe ansatz for the $15$-vertex model \cite{vieira2021solutions}.
These two cases are linked, as with certain choices of parameters, the $15$-vertex model can be mapped to a two-species PASEP with overtaking.
This suggests that condition \eref{pqpq} is generic in the context of exactly solvable models and not a special feature of the ansatz \eref{ansatz}.

In the subsequent calculations, it is convenient to use the following dimensionless parameters
\begin{eqnarray}
    x=q_2/p_2 \;,\\
   \beta_{12}=b_{12}/p_2 \;,\qquad \beta_{21}=b_{21}/p_2 \;.
\end{eqnarray}
Putting the ansatz \eref{ansatz} into the collision equations \eref{collpsi} gives
\numparts
\begin{eqnarray}
    [ {\cal D}_{ij} -(\alpha + x/\alpha )  z_j ] A^{11}_{ij} 
    + (i\leftrightarrow j)= 0 \;, \label{coll_nested_a} \\ [5pt]
    {[ {\cal D}_{ij}-(1 + x )  z_j ]} A^{22}_{ij} 
    + (i\leftrightarrow j) = 0 \;, \label{coll_nested_b} \\ [5pt]
    {[ ({\cal D}_{ij}-\kappa_2  z_j ) A^{12}_{ij}
    - \alpha \beta_{21} z_j A^{21}_{ij}]}
    + (i\leftrightarrow j) = 0 \;, \label{coll_nested_c} \\ [5pt]
    {[ ({\cal D}_{ij}-\kappa_1  z_j ) A^{21}_{ij}
    - \alpha^{-1}\beta_{12} z_j A^{12}_{ij}]}
    + (i\leftrightarrow j) = 0 \;,\label{coll_nested_d}
\end{eqnarray}
\endnumparts
where 
\begin{eqnarray}
    {\cal D}_{ij} = 1+xz_iz_j \;, \\
    \kappa _1 = 1+x/\alpha -\beta_{21} \;,\qquad \kappa _2 = \alpha +x-\beta_{12} \;,
\end{eqnarray}
and the notation $(i\leftrightarrow j)$ indicates a term similar to the explicitly written one but with the indices $i,j$ swapped.

The system of  equations \eref{coll_nested_a}--\eref{coll_nested_b} can be written compactly in vector form
\begin{eqnarray}
    [{\cal D}_{ij}1_{4\times 4}- z_j H]{\vec A}_{ij} + [{\cal D}_{ij}1_{4\times 4}- z_i H]{\vec A}_{ji} = 0 \;,
\end{eqnarray}
where
\begin{eqnarray}
    H = \left(
    \begin{array}{cccc}
        \alpha + x/\alpha & 0 & 0 & 0  \\
        0 & \kappa _2 & \alpha \beta_{21} & 0 \\
        0 & \beta_{12}/\alpha  & \kappa _1 & 0 \\
        0 & 0 & 0 & 1 + x  
    \end{array}
    \right) \;, \\[5pt]
    {\vec A}_{ij} = (A^{11}_{ij},A^{12}_{ij},A^{21}_{ij},A^{22}_{ij})^{\rm T} \;.
\end{eqnarray}
This can be formally rewritten as
\begin{eqnarray}
    {\vec A}_{ji} = 
    -\frac{{\cal D}_{ij}1_{4\times 4}- z_j H}{{\cal D}_{ij}1_{4\times 4}- z_i H}
    {\vec A}_{ij} \;, \label{collnestedfra}
\end{eqnarray}
where division is to be understood as multiplication by the matrix inverse from the left.
Performing this calculation explicitly, we get relations of the form
\begin{equation}
     A^{Q_1Q_2}_{ji} = 
    \sum\limits_{Q_1',Q_2'=1,2}S^{Q_1Q_2}_{Q_1'Q_2'}(z_i , z_j)
     A^{Q_1'Q_2'}_{ij} \;, \label{formfactors}
\end{equation}
in which $S^{Q_1Q_2}_{Q_1'Q_2'}(z_i , z_j)$ are called form factors.
Of these, only the following are non-zero
\numparts
\begin{eqnarray}
    S^{11}_{11}(z_i,z_j) = -\frac{{\cal D}_{ij}-(\alpha +x/\alpha)z_j}{{\cal D}_{ij}-(\alpha +x/\alpha)z_i} \;, \label{s1111}
    \\
    S^{22}_{22}(z_i,z_j) = -\frac{{\cal D}_{ij}-(1+x)z_j}{{\cal D}_{ij}-(1+x)z_i} \;, \label{s2222}\\
    S^{12}_{12}(z_i,z_j) = -\Phi_{ij}[{\cal D}_{ij}^2-{\cal D}_{ij}(\kappa_1 z_j + \kappa _2 z_i)+(\kappa _1 \kappa_2 - \beta_{12} \beta_{21})z_iz_j] \;, \label{s1212}\\
    S^{12}_{21}(z_i,z_j) = -\Phi_{ij}\alpha \beta_{21} {\cal D}_{ij}(z_i-z_j) \;, \label{s1221}\\
    S^{21}_{12}(z_i,z_j) = -\Phi_{ij}\alpha^{-1}\beta_{12} {\cal D}_{ij}(z_i-z_j) \;, \label{s2112}\\
    S^{21}_{21}(z_i,z_j) = -\Phi_{ij}[{\cal D}_{ij}^2-{\cal D}_{ij}(\kappa_2 z_j + \kappa _1 z_i)+(\kappa _1 \kappa_2 - \beta_{12} \beta_{21})z_iz_j] \label{s2121}\;,
\end{eqnarray}
\endnumparts
where
\begin{eqnarray}
   \Phi_{ij} = 
   [({\cal D}_{ij}-\kappa _1z_i)({\cal D}_{ij}-\kappa _2z_i) 
   -\beta_{12}\beta_{21}z_i^2]^{-1}\;.
\end{eqnarray}

\subsection{Bethe ansatz for deformed Markov operator}\label{BAdeformed}
Large deviations  in interacting particle systems can be derived by conditioning the dynamics  on a given, atypical value, of the observable under consideration.
The resulting evolution is governed by a deformation of the original Markov operator.
This method goes back to Donsker and Varadhan (see \cite{TouchetteReview,Chetrite} and references therein).
In the context of ASEP, this idea has been has been used systematically in conjunction with the BA, leading to exact results for the large deviation function of the current \cite{derrida1998exact,derrida1999bethe,prolhac2008current,Simon1,Simon2}.
We now outline this calculation for the two-species case, merely for completeness, since much of the detail is similar to the undeformed case, analyzed in the previous section.

We define observables $Y^{1}_t,Y^{2}_t,Y^{12}_t$, which count the number of processes $10\to 01\;,20\to 02\;,12\to 21$ minus the inverse processes respectively, up to time $t$.
Thus, for example, $Y^1_t+Y^{12}_t$ is the net total displacement of particles of species $1$ to the right.
We can split the Markov operator ${\cal M}$ into
\begin{eqnarray}
    {\cal M} = {\cal M}_0 + \sum\limits_{i=1,2,12} ({\cal M}_{i,+}+{\cal M}_{i,-}) \;,
\end{eqnarray}
where ${\cal M}_0$ contains the diagonal entries and ${\cal M}_{i,+(-)}$ contains the rates of the processes that increase (decrease) $Y^i_t$ by 1.
Let $P_t({\cal C},Y^1,Y^2,Y^{12})$ be the joint probability of configuration ${\cal C}$ with $Y^i_t=Y^i$ at time $t$.
Its generating function is defined as
\begin{equation}
    F_t ({\cal C},\gamma _1,\gamma _2, \gamma _{12}) = 
    \sum\limits _{Y^1,Y^2,Y^{12} = -\infty} ^{\infty}
    \rme ^{\gamma _1 Y^1 +\gamma _2 Y^2 +\gamma _{12} Y^{12}} P_t({\cal C},Y^1,Y^2,Y^{12}) \;.
\end{equation}
Note that by marginalizing over configurations, we obtain the moment generating function of $Y^{1}_t,Y^2_t,Y^{12}_t$,
\begin{equation}
    \sum\limits_{\cal C} F_t ({\cal C},\gamma _1,\gamma _2, \gamma _{12}) = 
    \langle \rme ^{\gamma _1 Y^1_t+\gamma _2 Y^2_t +\gamma _{12} Y^{12}_t}\rangle \;.
\end{equation}
A key property is that this satisfies a large deviation principle as $t\to \infty$,
\begin{equation}
    \langle \rme ^{\gamma _1 Y^1_t+\gamma _2 Y^2_t +\gamma _{12} Y^{12}_t}\rangle 
    \sim \rme ^{\mu (\gamma _1,\gamma _2,\gamma _{12})t}\;, \label{ldp}
\end{equation}
with some rate function $\mu (\gamma _1,\gamma _2,\gamma _{12})$.

From \eref{master}, we get the evolution equation
\begin{equation}
    \partial _t F_t ({\cal C},\gamma _1,\gamma _2, \gamma _{12}) = 
    \sum\limits_{{\cal C}'} {\cal M}_{\gamma }({\cal C},{\cal C}')F_t ({\cal C'},\gamma _1,\gamma _2, \gamma _{12}) \;,
\end{equation}
where ${\cal M}_\gamma $ is the deformed Markov operator
\begin{equation}
    {\cal M} _\gamma = {\cal M}_0 + \sum\limits_{i=1,2,12} 
    (\rme ^{\gamma _i}{\cal M}_{i,+}+\rme ^{-\gamma _i}{\cal M}_{i,-}) \;.
\end{equation}
The eigenvalue-eigenvector problem for ${\cal M}_\gamma $ then becomes
\begin{eqnarray}
    \fl
    \lambda (\gamma _1,\gamma_2,\gamma_{12})\psi ^{Q_1 \dots Q_M} (x_1 \dots x_M) = 
    -[M_1 (p_1+q_1) + M_2(p_2+q_2)]\psi ^{Q_1 \dots Q_M} (x_1\dots x_M) \nonumber \\
    \fl \qquad +\sum\limits_{j=1}^{M} 
    [p_{Q_j} \rme^{\gamma _{Q_j}}\psi ^{Q_1 \dots Q_M} (\dots  x_j -1\dots  )
    +q_{Q_{j}} \rme^{-\gamma _{Q_j}} \psi ^{Q_1 \dots Q_M} (\dots  x_j +1\dots )] \;.
\end{eqnarray}

As with the undeformed case, this equation is strictly valid only for configurations with no two particles being on adjacent sites, which is why $\gamma_{12}$ does not appear here.
The problem has to be supplemented with collision equations, which are given below.

A particularly important eigenvalue, is the one with the largest real part, which can be identified with the rate function $\mu (\gamma _1,\gamma _2,\gamma _{12})$ in \eref{ldp}.
From the Perron-Frobenius theorem it follows that this eigenvalue is unique.
In the limit $\gamma _1\to 0,\;\gamma_2\to 0,\;\gamma_{12}\to 0$, it converges to the unique $0$ eigenvalue of the (undeformed) Markov operator ${\cal M}$, whose eigenvector is the stationary measure.

The collision equations read
\begin{eqnarray}
    \fl -(p_{Q_j} + q_{Q_{j+1}})\psi^{\dots Q_jQ_{j+1}\dots}(\dots x_j,x_j+1 \dots) 
    +p_{Q_{j+1}}  \rme^{\gamma_{Q_{j+1}}} \psi^{\dots Q_jQ_{j+1}\dots}(\dots x_j,x_j \dots) 
    \nonumber \\ 
    \fl \qquad + q_{Q_j} \rme^{-\gamma_{Q_j}}\psi^{\dots Q_jQ_{j+1}\dots}(\dots x_j+1,x_j+1 \dots) 
    = -b_{Q_j Q_{j+1}}\psi^{\dots Q_jQ_{j+1}\dots}(\dots x_j,x_j+1 \dots)\nonumber \\
    \fl \qquad + b_{Q_{j+1}Q_j}\rme^{\gamma_{Q_{j+1}Q_j}}\psi^{\dots Q_{j+1}Q_{j}\dots}(\dots x_j,x_j+1 \dots)\;,
    \label{collpsidef}
\end{eqnarray}
where $\gamma_{21}=-\gamma_{12}$.
This problem is solved by the ansatz
\numparts
\begin{eqnarray}
    \psi ^{Q_1 \dots Q_M} (x_1 \dots x_M) =
    r_{Q_1} ^{x_1} \dots r_{Q_M} ^{x_M} \sum\limits _\sigma A^{Q_1 \dots Q_M}_\sigma
    z_{\sigma (1)}^{x_1} \dots  z_{\sigma (M)}^{x_M} \;, \label{ansatzdefa}\\
    r_1 = \alpha \rme ^{\gamma _1} \;,\;
    r_2 = \rme ^{\gamma _2}
    \;. \label{ansatzdefb}
\end{eqnarray}
\endnumparts

Plugging the ansatz \eref{ansatzdefa}-\eref{ansatzdefb} into the collision equations \eref{collpsidef} gives after some simplification an equation similar to \eref{collnestedfra}, with the only difference being that now
\begin{eqnarray}
    H = \left(
    \begin{array}{cccc}
        \alpha + x/\alpha & 0 & 0 & 0  \\
        0 & \kappa _2 & \alpha \beta_{21}\rme^{\bar \gamma} & 0 \\
        0 & \alpha ^{-1}\beta_{12} \rme^{-{\bar \gamma}} & \kappa _1 & 0 \\
        0 & 0 & 0 & 1 + x  
    \end{array}
    \right) \;,
\end{eqnarray}
where
\begin{equation}
    {\bar \gamma} = \gamma_1-\gamma_2-\gamma_{12}\;.
\end{equation}
Then repeating the calculation of the form factors, we get that most of them are the same as in the undeformed case, with only the following modifications
\numparts
\begin{eqnarray}
    S^{12}_{21}(z_i,z_j) = -\Phi_{ij}\alpha \beta_{21} \rme^{{\bar\gamma}} {\cal D}_{ij}(z_i-z_j) \;,\\
    S^{21}_{12}(z_i,z_j) = -\Phi_{ij}\alpha^{-1}\beta_{12}\rme^{-{\bar\gamma}} {\cal D}_{ij}(z_i-z_j) \;.
\end{eqnarray}
\endnumparts

\section{Yang-Baxter equations} \label{YBE}
\subsection{Derivation of YBE} \label{YBEderivation}

The Yang-Baxter relations are obtained by analyzing  collisions that involve 3 particles.
In this case, two sequences of binary collisions  can  result in the same permutation: $ijk\to ikj \to kij \to kji$ and $ijk \to jik \to jki \to kji$. 
In order for the nested coordinate BA to be consistent, the products of form factors obtained from these two sequences must be the same.
This requirement yields the YBE (see \cite{alcaraz1999exact} for a similar analysis).
We shall follow this procedure both for the regular and deformed cases, with the only difference being the explicit forms of the form factors.
Somewhat surprisingly, for the systems at hand, the results are insensitive to the deformation parameters.

The consistency condition gives some relations between the amplitudes $A^{Q_1Q_2Q_3}_{ijk}$ and $A^{Q_1'Q_2'Q_3'}_{kji}$, which for the two-species case is in principle an equation for $8\times 8$ matrices, called $R$-matrices.
The YBE can be formulated as an algebraic relation for the $R$-matrices.
In the context of driven diffusive systems this was done for instance in \cite{cantini2008algebraic}.
However, due to particle conservation, the relations can be simplified by dividing them into disjoint blocks with a fixed number of particles of each species.
It is not hard to verify that the $111$ and $222$ blocks are trivially satisfied for any choice of parameters.
For the $112$ block, we define
\begin{eqnarray}
    \fl
    \Xi (z_i,z_j)= 
    \left(
    \begin{array}{ccc}
        S^{11}_{11} & 0 & 0 \\
        0 & S^{12}_{12} & S^{12}_{21} \\
        0 & S^{21}_{12} & S^{21}_{21}
    \end{array}
    \right) _{(z_i,z_j)} \;, \;
    {\tilde\Xi} (z_i,z_j)= 
    \left(
    \begin{array}{ccc}
        S^{12}_{12} & S^{12}_{21} & 0 \\
        S^{21}_{12} & S^{21}_{21} & 0 \\
        0 & 0 & S^{11}_{11}
    \end{array}
    \right) _{(z_i,z_j)} \;.
\end{eqnarray}
Then the YBE are given by
\begin{equation}
    \Xi  (z_j,z_k) {\tilde\Xi} (z_i,z_k) \Xi  (z_i,z_j)- 
    {\tilde\Xi} (z_i,z_j) \Xi (z_i,z_k) {\tilde\Xi} (z_j,z_k)=0 \;.
    \label{yb}
\end{equation}
The YBE for the $122$ block can be derived similarly.

We remark that all the entries of the matrices $\Xi$ are rational functions of the Bethe roots.
Thus, when we combine the two matrix products in \eref{yb}, the result is a $3\times 3$ matrix whose entries are rational functions of the Bethe roots.
Their explicit forms are evidently very complicated but they can be analyzed with the help of a symbolic programming language, such as Mathematica.

\subsection{Solutions of YBE} \label{YBEsolution}
We wish to find conditions on the model parameters $(\alpha ,x, \beta _{12},\beta_{21} )$ that ensure that \eref{yb} are satisfied identically, for all values of the Bethe roots.

Firstly, we note that the numerators of the entries are polynomials in $z_i,z_j,z_k$, which must vanish identically.
We may therefore examine individual coefficients of $z_i,z_j,z_k$ in the numerators and require them to vanish.
This can be done in many ways, as there are many possible coefficients to look at.
Fortunately, taking only a few is sufficient to obtain conditions that are strong enough to satisfy the full set of equations.
One possible method is outlined in \ref{ybexample}.

We obtain three general solutions.
For the sake of conciseness, we do not list one-species models (which trivially form subsets of the models listed below and are known to be integrable), and models which are equivalent to the given ones up to permutation of species labels (including empty sites as a ``zeroth species").

We remark that all the following models satisfy the relation
\begin{equation}
    b_{12}b_{21}=p_2q_2 \;,
\end{equation}
which was not imposed {\it a priori}.
This shows that, the condition \eref{pqpq} cannot be circumvented by a relabelling of the species, which further supports the claim that it might be a necessary condition.

\subsubsection{Solution 1: TASEP case.}
The first solution is defined by the conditions
\begin{equation}
    x=0 \;, \qquad \beta_{21}=0 \;,
\end{equation}
which corresponds to the process
\begin{eqnarray*}
    10
    \overset{\alpha p_2}{\to}
    01 \quad;\quad
    20
    \overset{p_2}{\to}
    02 \quad;\quad
    12
    \overset{b_{12}}{\to}
    21 \;.
\end{eqnarray*}
\newline
This is a totally asymmetric process, with particles of species  1 hopping at rate $\alpha p_2$ and overtaking  particles  of species 2 at rate $b_{12}$.
The matrix product solution for the steady state was first given in \cite{derrida1993exact}.
The case of a single particle of species 2 (referred to as a defect) has been much studied in the literature \cite{mallick1996shocks,derrida1999bethe,cantini2008algebraic}.

\subsubsection{Solution 2: PASEP with first- and second-class particles.}
The second solution is defined by the conditions
\begin{equation}
    \alpha = 1 \;, \qquad \beta_{12} = 1 \;, \qquad \beta_{21} = x \; ,
\end{equation}
which corresponds to the process
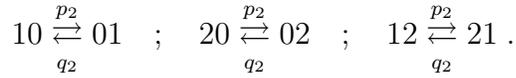
\begin{eqnarray*}
    10
    \underset{q_2}{\overset{p_2}{\rightleftarrows}}
    01 \quad;\quad
    20
    \underset{q_2}{\overset{p_2}{\rightleftarrows}}
    02 \quad ; \quad
    12
    \underset{q_2}{\overset{p_2}{\rightleftarrows}}
    21 \; .
\end{eqnarray*}
\newline
This is a two-species partially asymmetric process, with both species hopping at the same rates and the species 1 particles overtaking species 2 particles with their usual hopping rates. 
Since species 1 does not distinguish between species 2 and a vacancy, species 1 can be seen as having priority in the dynamics and is therefore referred to in the literature as first-class and species 2 as second-class.
This case has been studied in the literature as well, with the matrix product solution for the steady state given in \cite{derrida1993exact} and  BA equations derived in \cite{alcaraz1999exact,alcaraz2000exact}.
We remark that although the Bethe equations have been derived, the statistics of the current fluctuations have not yet been calculated.
\subsubsection{Solution 3: PASEP with a single first-class defect.}
The third solution is defined by the conditions
\begin{equation}
    \beta_{12} = \alpha \;, \qquad \beta_{21} = x/\alpha \;, \qquad M_1 = 1 \; ,
\end{equation}
which corresponds to the process
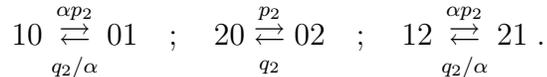
\begin{eqnarray*}
    10
    \underset{q_2/\alpha }{\overset{\alpha p_2}{\rightleftarrows}}
    01 \quad ; \quad
    20
    \underset{q_2}{\overset{p_2}{\rightleftarrows}}
    02 \quad ; \quad
    12
    \underset{q_2/\alpha }{\overset{\alpha p_2}{\rightleftarrows}}
    21  \;.
\end{eqnarray*}
\newline
This is a partially asymmetric process, with a single first-class defect particle with hopping rates $\alpha p_2,q_2/\alpha $ that overtakes the second-class particles with its usual hopping rates.
This model has not been analyzed in the literature until recently, when its steady state was solved using a matrix product approach \cite{lobaskin2022matrix}.

We remark that the constraint on particle number $M_1=1$ is unusual in the context of YBE.
It emerges here as the given conditions on the rates are sufficient to satisfy \eref{yb} for the $122$ block but not the $112$ block.
This means that the ansatz works for this general case only if there is exactly one particle of species 1 in the system.
To satisfy \eref{yb} for the $112$ block as well, it is necessary to impose either $\alpha =1$ or $\alpha=x$ or $x=0$.
In all these cases, this solution reduces down to solution 1 or 2.

\section{Derivation of Bethe equations for Solution 3} \label{newmodelBA}
As solution 3 has not appeared much in the literature, it presents a case of particular interest.
As mentioned previously, it was studied using a matrix product approach in \cite{lobaskin2022matrix}.
The phase diagram was shown to consist of localized phases, in which the defect particle has no macroscopic effect on the system, and a shock phase, in which the defect creates two extended regions with different bulk densities.
The density profiles and currents were also calculated.
The BA approach presented here complements these results, as it should allow one to calculate the fluctuations of the current as well.

To that end, we now derive the Bethe equations for the deformed eigenvalue problem from \sref{BAdeformed}.
As we have only one particle of species $1$, it is not necessary to use the full nested ansatz.
Instead, we can fix the particle indexing so that the species $1$ particle is always the leftmost one.
We can always do this using the boundary condition \eref{pbc}, which gives the following relation for amplitudes
\begin{eqnarray}
    A^{Q_1Q_2\dots }_{ij\dots} = \rme^{\gamma _{Q_1} L}z_i^L A^{Q_2\dots Q_1}_{j\dots i}\;.
\end{eqnarray}

Now that the order of species is always fixed to be $12\dots 2$, we can drop the upper indices without ambiguity.
Then we get for 22, 12 and 21 collisions respectively,
\numparts
\begin{eqnarray}
    A_{\dots ji\dots} =
    -\frac{{\cal D}_{ij}-(1+x)z_j}{{\cal D}_{ij}-(1+x)z_i} A_{\dots ij\dots}
    \;, \label{coll_a} \\ [5pt]
    {[ ({\cal D}_{ij} -xz_j )A_{ij\dots} 
    - x \rme^{\bar \gamma} z_i (\rme ^{\gamma _2}z_j)^L A _{i\dots j} ]} 
    + (i\leftrightarrow j) =0\;, \label{coll_b}\\ [5pt]
    {[ ({\cal D}_{ij}-z_i )(\rme ^{\gamma _2}z_j) ^L A_{i\dots j} 
    - \rme^{-{\bar\gamma}} z_j A_{ij\dots } ]} + (i\leftrightarrow j) =0\;. \label{coll_c}
\end{eqnarray}
\endnumparts
We will denote the form factor in \eref{coll_a} as $S^{22}_{22}(z_i,z_j)$ for conciseness.
Applying \eref{coll_a} $M_2-1$ times to \eref{coll_b} and \eref{coll_c} gives
\numparts
\begin{eqnarray}
     \left[ {\cal D}_{ij}-xz_j
     + x z_i S^{22}_{22}(z_i,z_j)E_j \right] 
     A_{ij\dots} + (i\leftrightarrow j) = 0 \;, \label{beq_0a}\\ [5pt]
    \left[ ({\cal D}_{ij}-z_i) S^{22}_{22}(z_i,z_j)E_j + z_j \right] A_{ij\dots} 
    + (i\leftrightarrow j)=0 \label{beq_0b}\;,
\end{eqnarray}
\endnumparts
where
\begin{equation}
    E_j = \rme^{\bar\gamma}(\rme^{\gamma _2}z_j) ^L 
    \prod\limits _{k=1}^{M}S^{22}_{22}(z_j,z_k) \;.
\end{equation}
In order for the eigenvector not to vanish, the system of equations \eref{beq_0a}-\eref{beq_0b} for $A_{ij\dots},A_{ji\dots}$ must be degenerate.
The equation resulting from this requirement can be simplified, although this procedure is algebraically somewhat cumbersome and not very illuminating.
We give the details in \ref{beqalgebra} for completeness.
In this calculation, it becomes convenient to consider the following transformation of the Bethe roots \cite{Kim,prolhac2008current},
\begin{equation}
    z_i = \frac{1-y_i}{1-xy_i} \;. \label{ztoy}
\end{equation}

Ultimately, one obtains a set of equations, called the Bethe equations, for the transformed Bethe roots $y_i$,
\begin{equation}
    \rme ^{{\bar\gamma}+L\gamma_2}
    \left( \frac{1-y_i}{1-xy_i}\right) ^L
    \prod\limits_{k=1}^{M}\frac{xy_i-y_k}{y_i-xy_k} 
    = -\frac{B-y_i}{Bx-y_i} \;,
    \label{beq_1}
\end{equation}
where $B$ is a constant.
The constant $B$ can be fixed by multiplying the Bethe equations for all $i$,
\begin{equation}
   \prod\limits_{i=1}^{M}\frac{B-y_i}{Bx-y_i}
    = \alpha ^{-L}\rme ^{-(L-M){\bar\gamma}-L\gamma_{12}} \;,
    \label{bfix}
\end{equation}
where we have implemented the periodic boundary condition \eref{pbc} as
\begin{equation}
    \alpha ^L \rme ^{L(\gamma_1+M_2\gamma_2)}z_1 ^L \dots z_M ^L = 1 \;. \label{periodicity}
\end{equation}

It is interesting to compare \eref{beq_1} to the corresponding equations for the well-known cases of a one-species PASEP \cite{prolhac2008current} and the TASEP with a defect \cite{derrida1999bethe,cantini2008algebraic}.
The one-species PASEP can also be simplified using a change of variable similar to \eref{ztoy}.
The resulting equation involves a product of fractions of linear differences of the transformed Bethe roots $y_i$, like \eref{beq_1}.

Meanwhile, the TASEP with a defect also features an additional constant, like $B$, that has to be fixed as part of solving the Bethe equations.
As will be shown in a forthcoming publication, this leads to one of the Bethe roots converging to a different point in the complex plane to the other roots in the $\gamma_1\to 0,\gamma_2\to0,\gamma_{12}\to0$ limit.

We now briefly comment on how one would proceed to solve the Bethe equations.
\eref{beq_1} is a system of polynomial equations of degree $L+M+1$ for each Bethe root $y_i$, in which the coefficients in turn depend on all the roots.
One would first need to solve \eref{beq_1} consistently with \eref{bfix}, then choose $M$ roots to satisfy \eref{periodicity} for some $L$-th root of unity.
This would then give an eigenvalue through \eref{lambda}.
In practice this is very challenging both analytically and numerically.
However, special functional methods have been developed that allow to calculate perturbative expansions of the largest eigenvalue in terms of $\gamma_1,\gamma_2,\gamma_{12} $ \cite{derrida1998exact,derrida1999bethe,prolhac2008current,prolhac2010tree}.
In a forthcoming publication, we will show how these methods can be applied to the problem at hand.

\section{Conclusion}

We have presented a nested coordinate BA that solves the two-species PASEP problem with overtaking, in both the deformed and simple (undeformed) cases, provided the condition \eref{pqpq} is satisfied.

The YBE were derived, which allowed us to assess the BA integrability of all models in this class. 
Although the derivation presented here is not entirely general, as a particular form of the nested coordinate Bethe wave function \eref{ansatz} was used, it is conjectured that the list of integrable models presented is exhaustive.

It was shown that the solutions fall into three cases,
namely a TASEP with two species that move at different rates, with one species overtaking the other with an arbitrary rate; a PASEP with two species moving at the same rate and one species treating the other as holes; and a PASEP with a single defect particle that moves with a different rate and does not distinguish between particles and holes.
These results are summarized in \tref{tab:model_list}.
Two of these cases are well-known in literature.
Namely, the TASEP with defect particles, and the second-class PASEP.

Solution 3 is a PASEP with a single first-class defect particle and overtaking.
This model has only been investigated recently, when its steady state was solved using the matrix product approach \cite{lobaskin2022matrix}.
It is peculiar among the solutions listed here in that it is only integrable for a single defect particle.
This follows from the fact that the YBE are satisfied for three-body collisions in which the particle species are $122$ and $222$ but not $112$.

\begin{table}[t]
    \centering
    \begin{tabular}{|c|c|c|}
        \hline
        Model & Parameter restrictions & Free parameters  \\
        \hline
        TASEP with defect species & $q_1=q_2=b_{21}=0$ & $p_1,\; b_{12}\;, M_1,\; M_2$ \\
        Second-class PASEP & $p_1=b_{12}=p_2,\; p_2=b_{21}=q_2$ & $q_2,\; M_1,\; M_2$ \\
        PASEP with one defect & $p_1 = b_{12} = \alpha p_2 ,\; q_1=b_{21}=q_2/\alpha,\; M_1=1$ &
        $q_2,\; \alpha,\; M_2$ \\
        \hline
    \end{tabular}
    \caption{All Bethe ansatz integrable sub-classes of the two-species partially asymmetric exclusion process with overtaking on a ring.
    The rate $p_2$ is not included as a free parameter as it can be seen as just setting the timescale.
    The particle numbers $M_1,M_2$ are included as free parameters to highlight that $M_1$ is fixed in the third case.}
    \label{tab:model_list}
\end{table}

It is interesting to note that that among the models examined here, all the models that are integrable in the normal (undeformed) case remain integrable after a deformation of the Markov operator.
This is {\it a priori} not obvious and it would be of interest to investigate whether integrability is preserved under deformation in other contexts as well.

Finally, we derived the Bethe equations for the deformed time evolution problem of solution 3 of the YBE.
This should allow one to calculate the long-time current fluctuations in this model.
This would complement the result for the mean current, which has already been calculated using a matrix product ansatz \cite{lobaskin2022matrix}.
The Bethe equations have similar features to the well-known one-species PASEP and two-species TASEP, which makes it amenable to the techniques developed for those cases.

Another possible direction for future work would be to consider some extensions of the traditional YBE, which have been proposed in the literature.
These include the dynamical YBE \cite{ABB96} and the braided YBE \cite{FR01,FR02}.
It would be interesting to investigate whether those generalizations can provide further insight in the context of asymmetric exclusion processes.

\section*{Acknowledgements}
IL acknowledges studentship funding from EPSRC under Grant No. EP/R513209/1.
The work of KM has been supported by the project RETENU ANR-20-CE40-0005-01 of the French National Research Agency (ANR).
For the purpose of open access, the authors have applied a Creative Commons Attribution (CC BY) licence to any Author Accepted Manuscript version arising from this submission.

\appendix 
\section{Derivation of solutions of Yang-Baxter equations} \label{ybexample}
\begin{figure}
    \centering
    \begin{tikzpicture}[style = {draw,->,thick}]
    \node [draw]{(2,1) in block 112, (1,2) in block 122}[sibling distance = 9cm,level distance = 2.5cm]
        child {node [draw]{(2,1) in block 122}[sibling distance = 3.5cm]
            child {node [draw,ultra thick]{S 1} edge from parent node [left,xshift=-0.3cm]{$x=0$}}
            child {node [draw]{(1,1) in block 112} [sibling distance = 2cm]
                child {node [draw, ultra thick]{S2} edge from parent node [left]{$\beta_{12}=1$ or $x$}}
                child {node [draw, ultra thick]{S1} edge from parent node [right,xshift=0.1cm,yshift=-0.2cm]{$x=0$}}
            edge from parent node[left,xshift=0.1cm]{$\alpha=1$}}
            child {node [draw]{block 122 solved}[sibling distance = 2.5cm]
                child {node [draw,ultra thick]{S 3}
                    edge from parent node [left,xshift=0.1cm,yshift=0.2cm]{$M_1=1$}}
                child {node [draw]{(1,2) in block 112}
                    child {node [draw,ultra thick]{S 1} edge from parent node [left,xshift=-0.3cm]{$x=0$}}
                    child {node [draw,ultra thick]{S 2} edge from parent node [right,xshift=0.3cm]{$\alpha=1$ or $x$}}
                    edge from parent node [right,xshift=0.3cm]{solve block 112}}
                edge from parent node [right,xshift=0.3cm]{$\beta_{12}=\alpha$}}
            edge from parent node [left,xshift=-0.5cm]{$\beta_{21}=x/\beta_{12}$}} 
        child {node [draw]{(1,1) in block 122}[sibling distance = 4cm]
            child {node [draw,ultra thick]{1-species PASEP} edge from parent node [left,xshift=-0.3cm]{$\alpha=1$}}
            child {node [draw,ultra thick]{S 1 (with $\beta_{12}=0$)} edge from parent node [right,xshift=0.3cm]{$x=0$}}
        edge from parent node [right,xshift=0.5cm]{$\beta_{12}=\beta_{21}=0$}};
    \end{tikzpicture}
    \caption{Representation of the procedure to obtain solutions to the Yang-Baxter equations described in \ref{ybexample}.
    Solutions are outlined with thick boundaries.
    S stands for solution.}
    \label{fig:tree}
\end{figure}
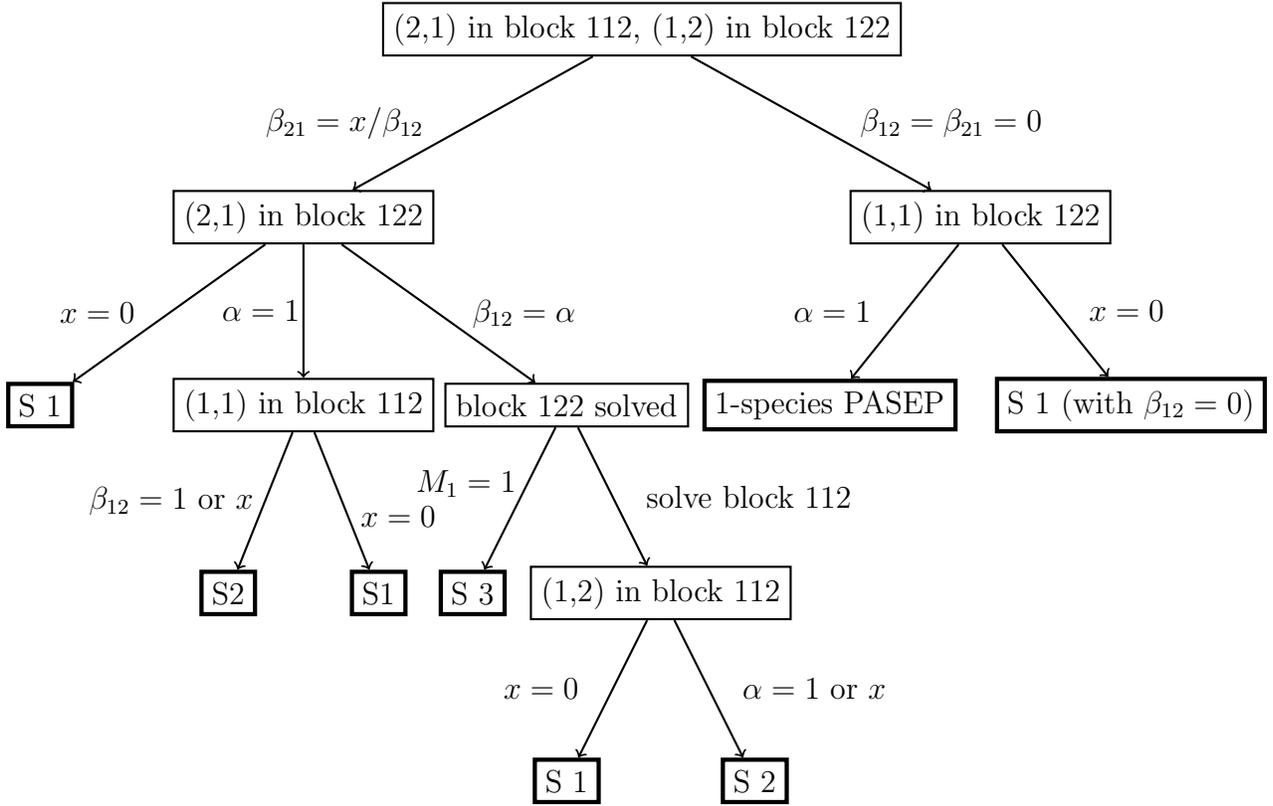

We wish to obtain conditions on the model parameters $\alpha ,x,\beta _{12},\beta_{21}$ such that \eref{yb} are satisfied for the 112 and 122 blocks identically for any values of $z_i,z_j,z_k$.
As explained previously, we can do this by looking at individual coefficients of $z_i^{n_i}z_j^{n_j}z_k^{n_k}$ for any powers $n_i,n_j,n_k$ and requiring them to vanish.
We perform the following calculations with the help of Mathematica to evaluate the needed expressions symbolically.
We show the expressions for the deformed case, as the undeformed limit can be obtained simply by setting $\gamma_1=\gamma_2=\gamma_{12}=0$.
The procedure presented here is shown schematically in \fref{fig:tree} for clarity.

First, we note that all entries have a factor of $(z_i-z_j)(z_i-z_k)(z_j-z_k)$, so we divide by this for simplicity.
This will be assumed to have been done in all subsequent expressions.
Then, looking at the constant term (i.e. taking $z_i\to0,z_j\to0,z_k\to0$), in the (2,1) entry in the 112 block and the (1,2) entry in the 122 block, we get respectively,
\begin{equation}
    \rme^{\gamma_2+\gamma_{12}}\alpha^4\beta_{12}(\beta_{12}\beta_{21}-x)=0\;,
    \qquad \rme^{\gamma_1}\alpha^4\beta_{21}(\beta_{12}\beta_{21}-x)=0 \;.
\end{equation}

We cannot set $\alpha=0$, as, due to \eref{r1r2}, this would make the ansatz vanish.
Hence we can either set $\beta_{21}=x/\beta_{12}$ or $\beta_{12}=\beta_{21}=0$.

\subsection*{Case 1: $\beta_{21}=x/\beta_{12}$}

With this choice, we now look at the constant term of the numerator in the (2,1) entry in the 122 block.
This gives us the additional condition
\begin{equation}
    \rme^{\gamma_2+\gamma_{12}}\alpha x(\alpha-1)(\alpha-\beta_{12})(1+\beta_{12})=0\;.
\end{equation}

As before, we cannot set $\alpha=0$.
This leaves us with three options: $x=0$ or $\alpha=1$ or $\beta_{12}=\alpha$.
The choice $x=0$ is sufficient to satisfy the YBE fully for the 112 and 122 blocks, giving solution 1.

With the choice $\alpha=1$, we now look at the coefficient of $z_i^0z_j^1z_k^1$ in the $(1,1)$ entry of the $122$ block.
This gives the additional condition
\begin{eqnarray}
    (1+x)x(1-\beta_{12})(x-\beta_{12})=0.
\end{eqnarray}
Setting $x=0$ gives solution 1.
Setting $\beta_{12}=1$ or $x$ is sufficient to satisfy the YBE for the $112$ and $122$ block.
Both options give solution 2 up to a relabelling of species.

The choice $\beta_{12}=\alpha$ is sufficient to satisfy the YBE for the 122 block.
The 112 block is not satisfied.
We can either limit ourselves to have only one particle of species 1, which gives solution 3, or try to satisfy the 112 block as well.
In the 112 block, looking at the constant term in the numerator of the (1,2) entry gives the condition
\begin{equation}
    \rme^{\gamma_1}\alpha^3x(\alpha-1)(\alpha^2+x)(\alpha-x)=0\;.
\end{equation}
If we wish to satisfy this, we must set either $x=0$ or $\alpha=1$ or $\alpha=x$.
All of those choices lead either to solution 1 or 2.

\subsection*{Case 2: $\beta_{12}=\beta_{21}=0$}

With this choice, we now look at the constant term in the numerator of the (1,1) entry of the 122 block.
This gives the further condition
\begin{equation}
    \alpha^3x(\alpha-1)=0\;.
\end{equation}

Setting $\alpha=1$ is enough to satisfy all YBE but it gives a essentially a one-species PASEP, which is not of interest here, as we are looking for two-species solutions.

Choosing $x=0$ gives a special case of solution 1 with $\beta_{12}=0$.

\section{Algebraic steps in derivation of Bethe equations for solution 3}\label{beqalgebra}

Requiring the system of equations \eref{beq_0a}-\eref{beq_0b} to be degenerate, we get
\begin{eqnarray}
    \fl \left[ {\cal D}_{ij}-xz_j + x z_i S^{22}_{22}(z_i,z_j)E_j \right]
     \left[ ({\cal D}_{ij}-z_j) S^{22}_{22}(z_j,z_i)E_i + z_i \right] 
     - (i\leftrightarrow j)=0 \;.
\end{eqnarray}

Expanding and simplifying this, we get
\begin{eqnarray}
    \fl ({\cal D}_{ij}-(1+x)z_j) S^{22}_{22}(z_j,z_i)E_i 
    - ({\cal D}_{ij}-(1+x)z_i) S^{22}_{22}(z_i,z_j)E_j \nonumber \\
    + (z_i-z_j)(1+xE_iE_j) = 0 \;.
\end{eqnarray}

Now plugging in the explicit forms of $S^{22}_{22}$, this becomes
\begin{equation}
    \fl -({\cal D}_{ij}-(1+x)z_i) E_i + ({\cal D}_{ij}-(1+x)z_j) E_j
    + (z_i-z_j)(1+xE_iE_j) = 0 \;.
\end{equation}

With the change of variable \eref{ztoy}, we get, after some simplification,
\begin{equation}
    (y_i-xy_j) E_i -(y_j-xy_i) E_j
    + (y_i-y_j)(1+xE_iE_j) = 0 \;,
\end{equation}
where now
\begin{equation}
     E_j = \rme^{{\bar\gamma} +L\gamma_2}
    \left(\frac{1-y_j}{1-xy_j}\right) ^L 
    \prod\limits _{k=1}^{M} \frac{xy_j-y_k}{y_j-xy_k} \;.
\end{equation}

The terms with index $i$ and $j$ can now be separated to give the following equation,
\begin{equation}
    y_i \frac{1+E_i}{1+xE_i} = y_j \frac{1+E_j}{1+xE_j} \;.
\end{equation}

As this procedure can be carried out for any two indices $i,j$, this means that this expression must equal some constant $B$ for all $i,j$.
Then equating the left-hand side with $B$ and rearranging to isolate $E_i$, we finally obtain \eref{beq_1}.
\section*{References}
\bibliographystyle{iopart-num.bst}
\bibliography{references}

\providecommand{\newblock}{}
\begin{thebibliography}{10}
\expandafter\ifx\csname url\endcsname\relax
  \def\url#1{{\tt #1}}\fi
\expandafter\ifx\csname urlprefix\endcsname\relax\def\urlprefix{URL }\fi
\providecommand{\eprint}[2][]{\url{#2}}

\bibitem{derrida2007non}
Derrida B 2007 {\em Journal of Statistical Mechanics: Theory and Experiment\/}
  {\bf 2007} P07023

\bibitem{blythe2007nonequilibrium}
Blythe R~A and Evans M~R 2007 {\em Journal of Physics A: Mathematical and
  Theoretical\/} {\bf 40} R333

\bibitem{chou2011non}
Chou T, Mallick K and Zia R~K 2011 {\em Reports on progress in physics\/} {\bf
  74} 116601

\bibitem{macdonald1968kinetics}
MacDonald C~T, Gibbs J~H and Pipkin A~C 1968 {\em Biopolymers: Original
  Research on Biomolecules\/} {\bf 6} 1--25

\bibitem{SNCM2018}
Szavits-Nossan J, Ciandrini L and Romano M~C 2018 {\em Phys. Rev. Lett.\/} {\bf
  120}(12) 128101

\bibitem{scott2019power}
Scott S and Szavits-Nossan J 2019 {\em Physical biology\/} {\bf 17} 015004

\bibitem{wolf1996traffic}
Wolf D~E, Schreckenberg M and Bachem A 1996 {\em Traffic and granular flow\/}
  (World Scientific)

\bibitem{cividini2017driven}
Cividini J, Mukamel D and Posch H~A 2017 {\em Physical Review E\/} {\bf 95}
  012110

\bibitem{dehp1993}
Derrida B, Evans M~R, Hakim V and Pasquier V 1993 {\em Journal of Physics A:
  Mathematical and General\/} {\bf 26} 1493

\bibitem{Alexander}
Alexander S and Holstein T 1978 {\em Phys. Rev. B\/} {\bf 18}(1) 301--302

\bibitem{Dhar}
Dhar D 1987 {\em Phase transitions\/} {\bf 1} 51

\bibitem{gwa1992bethe}
Gwa L~H and Spohn H 1992 {\em Physical Review A\/} {\bf 46} 844

\bibitem{golinelli2006asymmetric}
Golinelli O and Mallick K 2006 {\em Journal of Physics A: Mathematical and
  General\/} {\bf 39} 12679

\bibitem{bethe1931theorie}
Bethe H 1931 {\em Zeitschrift f{\"u}r Physik\/} {\bf 71} 205--226

\bibitem{baxter2016exactly}
Baxter R~J 2016 {\em Exactly solved models in statistical mechanics\/}
  (Elsevier)

\bibitem{schutz1997exact}
Sch{\"u}tz G~M 1997 {\em Journal of statistical physics\/} {\bf 88} 427--445

\bibitem{sasamoto1998one}
Sasamoto T and Wadati M 1998 {\em Physical Review E\/} {\bf 58} 4181

\bibitem{de2005bethe}
De~Gier J and Essler F~H 2005 {\em Physical review letters\/} {\bf 95} 240601

\bibitem{simon2009construction}
Simon D 2009 {\em Journal of Statistical Mechanics: Theory and Experiment\/}
  {\bf 2009} P07017

\bibitem{Kim}
Kim D 1995 {\em Phys. Rev. E\/} {\bf 52}(4) 3512--3524

\bibitem{alcaraz1999exact}
Alcaraz F and Bariev R 1999 {\em Physical Review E\/} {\bf 60} 79

\bibitem{alcaraz2000exact}
Alcaraz F and Bariev R 2000 {\em Brazilian Journal of Physics\/} {\bf 30}
  655--666

\bibitem{de2008slowest}
de~Gier J and Essler F~H 2008 {\em Journal of Physics A: Mathematical and
  Theoretical\/} {\bf 41} 485002

\bibitem{prolhac2008current}
Prolhac S and Mallick K 2008 {\em Journal of Physics A: Mathematical and
  Theoretical\/} {\bf 41} 175002

\bibitem{prolhac2010tree}
Prolhac S 2010 {\em Journal of Physics A: Mathematical and Theoretical\/} {\bf
  43} 105002

\bibitem{derrida1999bethe}
Derrida B and Evans M~R 1999 {\em Journal of Physics A: Mathematical and
  General\/} {\bf 32} 4833

\bibitem{cantini2008algebraic}
Cantini L 2008 {\em Journal of Physics A: Mathematical and Theoretical\/} {\bf
  41} 095001

\bibitem{derrida1998exact}
Derrida B and Lebowitz J~L 1998 {\em Physical review letters\/} {\bf 80} 209

\bibitem{jimbo1990yang}
Jimbo M 1990 {\em Yang-Baxter equation in integrable systems\/} vol~10 (World
  Scientific)

\bibitem{kulish1982solutions}
Kulish P and Sklyanin E 1982 {\em Journal of Soviet Mathematics\/} {\bf 19}
  1596--1620

\bibitem{vieira2021solutions}
Vieira R~S and Lima-Santos A 2021 {\em Journal of Statistical Mechanics: Theory
  and Experiment\/} {\bf 2021} 053103

\bibitem{Schutz}
Sch\"{u}tz G~M 2001 {\em Exactly Solvable Models for Many-Body Systems Far from
  Equilibrium, in Phase Transitions and Critical Phenomena, C. Domb and J. L.
  Lebowitz (Eds)\/} vol~19 (Academic Press)

\bibitem{Crampe2}
Cramp\'{e} N, Mallick K, Ragoucy E and Vanicat M 2015 {\em J. Phys. A:
  Mathematical and Theoretical\/} {\bf 48} 175002

\bibitem{crampe2014integrable}
Cramp\'{e} N, Ragoucy E and Vanicat M 2014 {\em Journal of Statistical
  Mechanics: Theory and Experiment\/} {\bf 2014} P11032

\bibitem{lobaskin2022matrix}
Lobaskin I, Evans M~R and Mallick K 2022 {\em Journal of Physics A:
  Mathematical and Theoretical\/} {\bf 55} 205002

\bibitem{derrida1993exact}
Derrida B, Janowsky S~A, Lebowitz J~L and Speer E~R 1993 {\em Journal of
  statistical physics\/} {\bf 73} 813--842

\bibitem{mallick1996shocks}
Mallick K 1996 {\em Journal of Physics A: Mathematical and General\/} {\bf 29}
  5375

\bibitem{arndt1998spontaneous}
Arndt P~F, Heinzel T and Rittenberg V 1998 {\em Journal of Physics A:
  Mathematical and General\/} {\bf 31} L45

\bibitem{rajewsky2000spatial}
Rajewsky N, Sasamoto T and Speer E 2000 {\em Physica A: Statistical Mechanics
  and its Applications\/} {\bf 279} 123--142

\bibitem{personal}
Cantini L and Sasamoto T 2022 {\em Personal communication\/}

\bibitem{TouchetteReview}
Touchette H 2009 {\em Physics Reports\/} {\bf 478} 1--69 ISSN 0370-1573

\bibitem{Chetrite}
Chetrite R and Touchette H 2015 {\em Ann. Henri Poincare\/} {\bf 16} 2005

\bibitem{Simon1}
Simon D 2009 {\em J. Stat. Mech: Theor and Exp.\/} {\bf 2009} P07017

\bibitem{Simon2}
Popkov V, Sch\"{u}tz G~M and Simon D 2010 {\em J. Stat. Mech: Theor and Exp.\/}
  {\bf 2010} P10007

\bibitem{ABB96}
Avan J, Babelon O and Billey E 1996 {\em Communications in mathematical
  physics\/} {\bf 178} 281--299

\bibitem{FR01}
Fioravanti D and Rossi M 2001 {\em Journal of Physics A: Mathematical and
  General\/} {\bf 34} L567

\bibitem{FR02}
Fioravanti D and Rossi M 2002 {\em Journal of Physics A: Mathematical and
  General\/} {\bf 35} 3647

\end{thebibliography}
\end{document}